\input{aipcheck}

\documentclass[
    ,final            
  ]
  {aipproc}

\layoutstyle{6x9}

\newcommand{\ket}[1]{| {#1} \rangle}
\newcommand{\bra}[1]{\langle {#1} |}

\usepackage{amsmath}

\begin{document}

\title{\bf 
Linear-response calculation in
 the time-dependent density functional theory}


\author{Takashi~Nakatsukasa}{
  address={RIKEN Nishina Center, Wako-shi 351-0198, Japan}
  ,altaddress={Center for Computational Sciences, University of Tsukuba,
           Tsukuba 305-8571, Japan}
}

\author{Tsunenori~Inakura}{
  address={RIKEN Nishina Center, Wako-shi 351-0198, Japan}
}

\author{Paolo~Avogadro}{
  address={Departimento di Fisica, Universit\`a degli Studi di Milano, via Celoria 16, 20133 Milan, Italy}
  ,altaddress={RIKEN Nishina Center, Wako-shi 351-0198, Japan}
}

\author{Shuichiro~Ebata}{
  address={Center for Nuclear Study, University of Tokyo, Bunkyo-ku, 113-0033, Japan}
  ,altaddress={RIKEN Nishina Center, Wako-shi 351-0198, Japan}
}

\author{Koichi~Sato}{
  address={RIKEN Nishina Center, Wako-shi 351-0198, Japan}
}
\author{Kazuhiro Yabana}{
  address={Center for Computational Sciences, University of Tsukuba,
           Tsukuba 305-8571, Japan}
  ,altaddress={RIKEN Nishina Center, Wako-shi 351-0198, Japan}
}

\begin{abstract}
Linear response calculations based on the time-dependent
density-functional theory are presented.
Especially, we report results of the
finite amplitude method which we have recently proposed as
an alternative and feasible approach to
the (quasiparticle-)random-phase approximation.
Calculated properties of the giant resonances and low-energy
$E1$ modes are discussed.
We found a universal linear correlation between the low-energy $E1$ strength
and the neutron skin thickness.
\end{abstract}

\maketitle


\section{Introduction}

Nuclei are created in stars via nuclear reactions
where the temperatures are extremely high.
Understanding of astrophysical phenomena generally requires 
many subfields of physics.
Among them, nuclear physics plays key roles in the generation of
elements, the evolution of the stars, the energy production in the universe,
etc.
The lightest elements up to helium were produced in the Big Bang.
The other heavier elements are generated by many kinds of nuclear reactions
in stars.
Especially, in stellar explosions, there are thousands of reactions
supposed to take place producing a variety of radioactive isotopes.
However, to date, their reaction rates have not been hardly determined
experimentally.

To overcome these experimental difficulties, reliable theoretical
information is necessary.
Traditional ab-initio methods start from a nucleon-nucleon potential which
describes nucleon-nucleon scattering data.
However, since the nuclear systems are strongly correlated because of
a repulsive core in the potential,
their description requires highly sophisticated many-body methods,
such as the quantum monte carlo method \cite{PW01}.
To describe the nucleus in a quantitative way,
they must employ an additional three-body force.
These ab-initio methods are so involved that, even at present,
their investigations have
been limited to very light nuclei and to the homogeneous nuclear matter
\cite{HP00}.

Under these circumstances, it is highly demanded to establish
a universal theoretical approach which is able to describe properties of
all species of nucleus.
The nuclear density functional theory \cite{BHR03}
 is the most promising candidate among many nuclear models.
It was referred to as the self-consistent mean-field model, but its
concept is analogous to the density functional theory for many-electron
systems \cite{PY89,DG90}.
The functionals have typically about ten parameters which are adjusted by
extensive fits to nuclear structure data.
The most prominent feature of the model is that
a single energy functional enables us to
quantitatively describe almost all nuclei
in the nuclear chart and infinite nuclear/neutron matter as well.

The nuclear density functional models have been extensively used
since 1970s \cite{VB72}.
In the beginning, there were several restrictions related to
symmetries of the wave functions, which limited applications of the model.
In 1990s, significant computational advances together with vast amount of
new spectroscopic data obtained with large $gamma$-ray arrays
changed the situation.
Using the cranking prescription,
the density functional models were becoming a standard tool to
study rotational bands in heavy nuclei \cite{AFN90}.
In particular, the models were very successful in studies of
superdeformed rotational bands at high spin in $A=150$ \cite{AFN90,NMMN95}
and $A=190$ regions \cite{NMMS96}.
It should be emphasized that the model parameters were never adjusted
to these bands at all.
In 2000s, systematic calculations of nuclear ground-state properties
were performed, to predict properties of nuclei far from the stability
line \cite{LPT03}.
At the same time, experimental developments greatly increased
our knowledge on radioactive isotopes, then, had an impact on
the energy density functionals.
The nuclear density functional methods has now reached a point
where one need to introduce ``correlations beyond the Kohn-Sham scheme''
to further improve the quality of the description.

The density functional theory is designed for the description of
ground-state properties.
Its extension to a time-dependent density functional theory is
formally straightforward, which is also analogous to the
time-dependent mean-field theory.
The density matrix $\rho(t)$ obeys the following equation,
\begin{equation}
i\ \hbar\frac{\partial}{\partial t} \rho(t) = \left[ h(t), \rho(t) \right] ,
\end{equation}
where $h(t)$ is the Kohn-Sham (mean-field) Hamiltonian which is
a functional of $\rho(t)$.
In the case of open-shell superfluid nuclei with pairing,
the density $\rho(t)$ and the Hamiltonian $h(t)$
should be generalized to those with double size of matrices, $R(t)$ and $H(t)$,
respectively \cite{BR86}.
Namely, $R(t)$ contains not only the normal density $\rho(t)$ but also the
pairing tensor $\kappa(t)$.
Similarly, $H(t)$ has the pair potential $\Delta(t)$ in addition to $h(t)$.
It is not yet fully understood how we can compute excitation properties
using the time-dependent density-functional theory, except for
a few limiting cases.
One such case is the low-energy regime of the nuclear dynamics,
such as surface vibrations and shape fluctuations, 
which can be reached by the adiabatic limit of
the time-dependent density-functional theory.
It is closely related to the microscopic derivation of the Bohr
model \cite{RS80}.
Recently, we have performed the numerical calculations of
the large-amplitude quadrupole dynamics in various isotopes
\cite{HSNMM10,HSYNMM11,YH11,SH11}.

Another important limiting case is the small-amplitude limit which provides
us with a powerful tool to study linear response in nuclei.
It is known as the random-phase approximation (RPA) or
the quasiparticle-random-phase approximation (QRPA) in nuclear physics,
which accesses the regime of giant resonances.
However, since the calculation of the residual interaction is
tedious for the realistic energy functionals, it has been common to
ignore some terms in practice and to sacrifice the full
self-consistency.
To overcome these difficulties and
facilitate an implementation of the full self-consistency,
we employ two different methodologies;
one is based on the real-time method (RTM) \cite{NY05}
and the other is the finite amplitude method (FAM) \cite{NIY07}.
The RTM has an advantage that it does not require the
calculation of the complex residual interactions, although
it has a limitation in the achieved
energy resolution inversely proportional to the time duration $T$.
Our recent applications of the RTM are based on the canonical-basis
framework \cite{Eba10} which is able to take into account
dynamical pairing effects in nuclei.
This is presented in another contribution to this volume \cite{ENI12}.
In contrast, the FAM, which is complementary to the RTM,
is a method of calculating the matrix elements of the residual
field, $\delta h = \delta h/\delta \rho \cdot \delta \rho$,
using the finite difference.
This does not require excessive programming but
can be done by employing the program of the static density-functional
calculation.
We have performed systematic and fully self-consistent RPA
calculations of photoabsorption cross sections for wide mass region ($A\le
100$), for both spherical and deformed nuclei.
In this report, we show results of these symmetry-unrestricted
FAM calculations.

\section{Finite Amplitude Method}

In this section, we recapitulate the FAM.

\subsection{FAM for RPA}

First, we discuss the case without the pairing correlations.
In this case, the FAM leads to residual fields appearing in the RPA.
For more details, readers are referred to Ref.~\cite{NIY07}.

The linear-response RPA equation to a weak external field with a fixed
frequency, $V_\mathrm{ext}(\omega)$, can be expressed in terms of
the forward and backward amplitudes, $\ket{X_i(\omega)}$ and
$\bra{Y_i(\omega)}$.
\begin{eqnarray}
\omega \, | X_i(\omega) \rangle = \left( h_0 - \epsilon_i \right) | X_i(\omega) \rangle
          + \hat{P} \left\{ V_\mathrm{ext}(\omega) + \delta h(\omega) \right\} | \phi_i \rangle  , \label{RPAeqX}\\
- \omega \, \langle Y_i(\omega) | = \langle Y_i(\omega) | \left( h_0 - \epsilon_i \right)
          + \langle \phi_i | \left\{ V_\mathrm{ext}(\omega) + \delta h(\omega) \right\} \hat{P}  , \label{RPAeqY}
\end{eqnarray}
where the subscript $i$ indicates the occupied orbitals
($i=1,2,\cdots,A$) and the operator $\hat{P}$ denotes the projector
onto the particle space, $\hat{P} =  1 -  \sum_i | \phi_i \rangle
\langle \phi_i |$.
Usually, the induced residual field $\delta h(\omega)$ is
expanded to the first order with respect to
$| X_i(\omega)\rangle$ and
$| Y_i(\omega)\rangle$.
This leads to the well-known matrix form of
the linear-response equation and calculation of these matrix
elements is most time-consuming in practice. Instead, we utilize the
fact that the linearization is numerically achieved for
$\delta h(\omega) = h[\rho_0 + \delta\rho(\omega)] - h_0$ within the
linear approximation. In order to perform this numerical
differentiation in the program, we use a small trick in the
calculation of the single-particle Hamiltonian $h[\rho]$.

First, we should notice that the $\delta h(\omega)$ depends only on
the forward "ket" amplitudes $\ket{X_i(\omega)}$ and
backward "bra" ones $\bra{Y_i(\omega)}$.
In other words, it is independent of
bras $\bra{X_i(\omega)}$ and kets $\ket{Y_i(\omega)}$.
This is related to the fact that the transition density $\delta\rho(\omega)$
depends only on $\ket{X_i(\omega)}$ and $\bra{Y_i(\omega)}$.
\begin{equation}
\delta\rho(\omega) = \sum_i \left\{ \ket{X_i(\omega)}\bra{\phi_i}
                                  + \ket{\phi_i}\bra{Y_i(\omega)}
                            \right\} .
\end{equation}
Then, we can calculate the residual fields in a following manner \cite{NIY07}:
\begin{equation}
\delta h(\omega) = \frac{1}{\eta} 
\left( h\left[ {\rho}_\eta \right] -
 h_0 \right) ,
\label{FAM}
\end{equation}
where $\eta$ is a small real parameter to realize the
linear approximation.
$\rho_\eta$ are defined by
\begin{equation}
{\rho}_\eta \equiv \sum_i \left\{
(\ket{\phi_i}+\eta\ket{X_i(\omega)})(\bra{\phi_i}+\eta\bra{Y_i(\omega)})
\right\} .
\end{equation}
Once $\ket{X_i(\omega)}$ and $\bra{Y_i(\omega)}$ are given,
the calculation of $h[\rho_\eta]$ is an easy task.
This does not require complicated programming, but only needs a small
modification in the calculation of $h[\rho]$.
Of course, eventually, we need to solve Eqs.~(\ref{RPAeqX}) and (\ref{RPAeqY})
to determine the forward and backward amplitudes.
We use an iterative algorithm to solve this problem.
Namely, we start from initial amplitudes $\ket{X_i^{(0)}}$ and
$\bra{Y_i^{(0)}}$, then update them in every iteration,
($\ket{X_i^{(n)}},\bra{Y_i^{(n)}}) \rightarrow
(\ket{X_i^{(n+1)}},\bra{Y_i^{(n+1)}})$,
until the convergence.
In each step, we calculate $\delta h(\omega)$ using the FAM
as Eq.~(\ref{FAM}).

\subsection{FAM for QRPA}

The FAM in the previous section can be extended to superfluid nuclei,
namely, to the QRPA with the Bogoliubov extension of the mean fields.
For more details, readers are referred to Ref.~\cite{AN11}.

A self-consistent solution of static Kohn-Sham-Bogoliubov problems
determines the ground-state densities $(\rho_0,\kappa_0)$
and the ground-state Hamiltonians $(h_0,\Delta_0)$.
They are given in terms of the quasiparticle wave functions,
$(U_\mu, V_\mu)$.
Then, following the same argument in the previous section,
we can derive equations for the residual fields, $\delta h(\omega)$
and $\delta\Delta(\omega)$ as follows \cite{AN11}:
\begin{equation}
\label{FAM_QRPA}
\begin{split}
\delta h(\omega) &= \frac{1}{\eta} \left(
h[\rho_\eta,\kappa_\eta] - h_0 \right) , \\
\delta \Delta(\omega) &= \frac{1}{\eta} \left(
\Delta[\rho_\eta,\kappa_\eta] - \Delta_0 \right) ,
\end{split}
\end{equation}
where the density and pairing tensor $(\rho_\eta, \kappa_\eta)$ are defined by
\begin{eqnarray}
\rho_\eta &=& ( V^* + \eta U X ) (V + \eta U^* Y )^T ,\\
\kappa_\eta &=& ( V^* + \eta U X ) ( U + \eta V^* Y )^T .
\end{eqnarray}
Here, the forward and backward amplitudes $(X_{\mu\nu},Y_{\mu\nu})$
have subscripts $\mu\nu$ to specify two-quasiparticles.
On the other hand, the subscripts of $(U_{k\mu},V_{k\nu})$ indicate
a basis of the single-particle space ($k$) and the quasiparticle ($\mu$).
Again, utilizing an iterative algorithm for solution of the QRPA equation,
we can solve the QRPA linear-response equation
without explicitly calculating the residual interactions.

\subsection{Numerical results}

\subsubsection{Development of FAM computer programs}
We have developed the computer codes of the FAM in several
representations.
The FAM-RPA is available in the three-dimensional (3D) grid representation
\cite{INY09,INY11}.
This provides a completely symmetry unrestricted RPA calculation.
All the single-particle wave functions and RPA amplitudes
are represented by these grid points:
\begin{equation}
\left\{ \phi_i(\vec{r}_k,\sigma), X_i(\omega;\vec{r}_k,\sigma),
Y_i(\omega;\vec{r}_k,\sigma)\right\}_{k=1,\cdots,N_{\rm grid}}
^{i=1,\cdots,A;\ \sigma={\rm up, down}} .
\end{equation}
Since the results are not sensitive to mesh spacing in a region
outside of the interacting region, the adaptive grid representation
is used to reduce the number of grid points $N_{\rm grid}$ \cite{NY05}.
In the followings, we show results obtained with this code.

The code for FAM-QRPA was also developed, but still with some
symmetry restrictions at present.
In Ref. \cite{AN11},
we developed a code in the radial coordinate representation based on
the spherical static program {\sc hfbrad} \cite{DFT84},
which assumes the spherical symmetry in the ground state.
Lately, another FAM-QRPA code has been developed
in the harmonic oscillator representation \cite{Sto11},
which is based on the program {\sc hfbtho} with the axial symmetry \cite{Sto03}.

In the present applications, we use complex energy,
$\omega=E+i\Gamma/2$, with $\Gamma=1$ MeV,
which introduces an artificial damping width.
This smoothing is necessary in two reasons:
To obtain smooth strength functions, and to speed up the convergence
for the iterative procedure.
However, the FAM formulae, Eqs. (\ref{FAM}) and (\ref{FAM_QRPA}) themselves,
are independent from the smoothing parameter $\Gamma$.
In fact, we can use the FAM for explicit construction of the (Q)RPA matrix
and calculate the discrete normal modes.
This will be reported elsewhere \cite{AN12}.

\begin{figure}
  \includegraphics[height=.3\textheight]{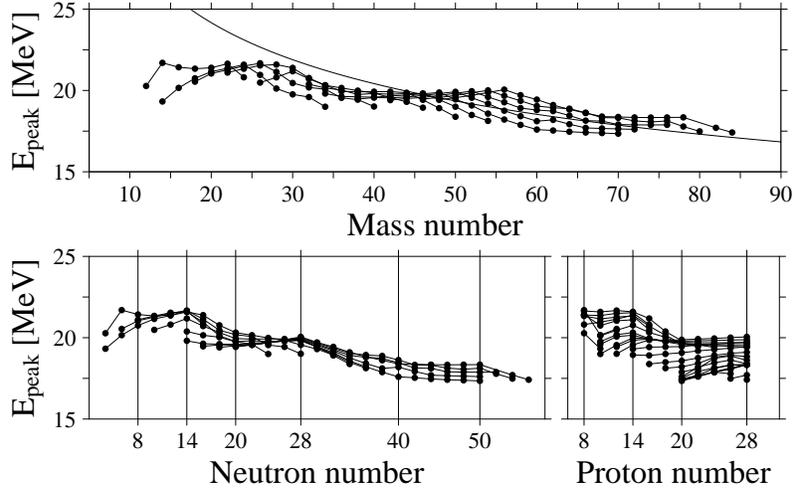}
  \caption{Calculated GDR peak energies as functions of mass number (top),
neutron number (bottom left), and proton number (bottom right).
The isotopic chains are connected by lines for the top and bottom left panels,
while the isotonic chains are shown in the bottom right.}
  \label{peak_position}
\end{figure}
\subsubsection{Giant dipole resonances}

We have carried out the systematic calculation of electric dipole
response in the FAM-RPA with the 3D grid representation.
We show evolution of peak energies of the giant dipole resonances (GDR)
as functions of mass number, neutron number, and proton
number in Fig. \ref{peak_position}. The GDR peak position is
estimated by the average energy
\begin{equation}
\label{m_k}
E_\mathrm{peak} = \frac{m_1(\omega_{\rm max})}{m_0(\omega_{\rm max})}
\,,\quad
m_k(\omega_{\rm max}) =
 \int_0^{\omega_{\rm max}} \mathrm{d}\omega \, \omega^k S(\omega; D_{E1})
\end{equation}
where $S(\omega;D_{E1})=\sum_n |\langle n|D_{E1}|0\rangle|^2
       \delta(\omega-E_n)$.
The maximum energy is $\omega_{\rm max}\approx 40$ MeV.
The $E1$ operator is defined with the recoil charges for protons,
$Ne/A$ and for neutrons, $-Ze/A$.
In deformed nuclei, since the peak energy depends on
direction of the $E1$ operator ($x$, $y$, and $z$),
their averaged value is adopted in Fig.~\ref{peak_position}.
The upper panel shows the GDR peak energies from oxygen to
nickel, as a function of mass number. In the medium-mass region,
the peak energies approximately follow the empirical low,
$21 A^{-1/3} + 31 A^{-1/6}$ MeV,
denoted by the solid curve. However, in each isotopic chain,
the peak energies in stable nuclei are the highest, while they are
decreasing as leaving from the stability line.
In addition, we can see some kind of shell effects.
This can be more clearly seen in
the lower panels of Fig. \ref{peak_position}, in which
the peak energies are plotted as functions of neutron and proton numbers.
There are cusps
at $N=$14 and 28 corresponding to the subshell closure of
$1d_{5/2}$ and $1f_{7/2}$ orbitals.
This may be attributed to the emergence of low-energy pygmy peaks
beyond these neutron numbers
(see the discussion in the next section).
The proton shell effects seem not to be as significant as those of
neutrons (see the bottom right panel of Fig.~\ref{peak_position}).

\begin{figure}
  \includegraphics[width=\textwidth]{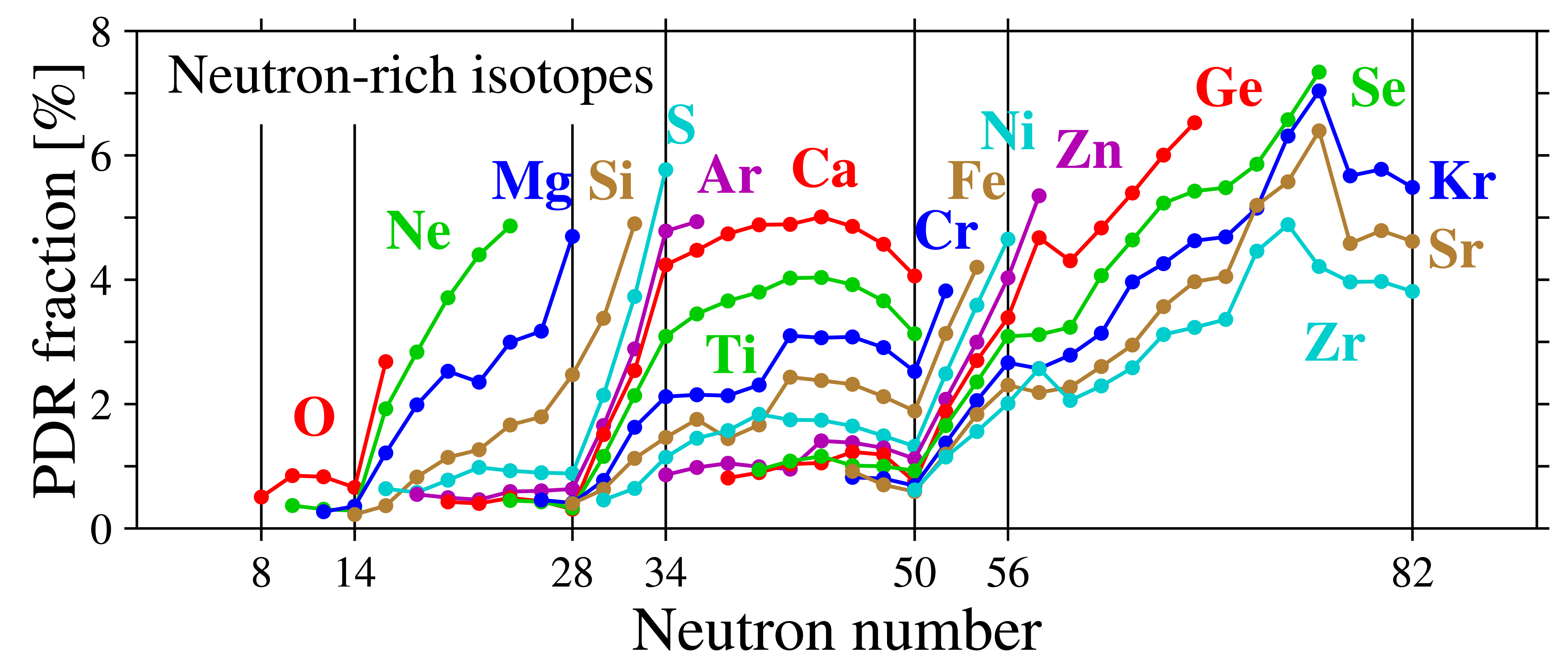}
  \caption{Calculated PDR strength fraction as a function of neutron number.
}
  \label{PDR_Ndep}
\end{figure}
\subsubsection{Low-energy $E1$ strength}

Now, let us move to the low-energy part of the $E1$ strength
distribution.
In some nuclei, there appear small peaks in the $E1$ strength
distribution, well separated from the main GDR peak,
which are often called ``pygmy dipole resonance'' (PDR).
In contrast to the main peak of GDR, the PDR strength distribution
is sensitive to
nuclear properties at nuclear surface and at low density.
Thus, its property may provide us with useful constraints on
the energy density functional,
to identify the equation of state (EOS) of
the nuclear and neutron matters.
For instance, the neutron skin thickness is known to be well
correlated with the slope of the neutron-matter EOS \cite{Bro00}.
Thus, if the neutron skin thickness has a strong correlation with
the low-energy $E1$ strength, we may pin down the EOS property
by observing the PDR in experiments.

First, we define the PDR strength fraction as
\begin{equation}
f_{\rm PDR} \equiv m_1(\omega_c)/m_1(\infty) ,
\end{equation}
where $m_1(\omega)$ is given in Eq. (\ref{m_k}) and we adopt
$\omega_c=10$ MeV.
In Fig.~\ref{PDR_Ndep},
we show the neutron-number dependence of $f_{\rm PDR}$.
It indicates a strong shell effect.
Namely, there are clear kinks at $N=14$, 28, and 50.
Let us concentrate our discussion on the kinks at $N=28$.
The PDR fractions suddenly increase at $N=28\rightarrow 30$ and
continue to increase till $N=34$ where the neutron $2p$ shell are filled.
Beyond $N=34$,
the PDR fractions are roughly constant for $34<N\leq 50$, in which
the neutrons are filling high-$\ell$ orbits of $f_{5/2}$ and $g_{9/2}$.
Beyond $N=50$, the neutrons start filling 
$2d_{5/2}$ orbits, then the $f_{\rm PDR}$ again shows a sudden increase.
These behaviors strongly suggest that
the spatially extended nature of the low-$\ell$
neutron orbits near the Fermi level plays a primary role for the
emergence and growth of the PDR.
We have also observed that the deformation tends to increase the PDR
strength, especially in the region $N>56$.
More detailed analysis can be found in our recent paper \cite{INY11}.

\begin{figure}
  \includegraphics[width=0.72\textwidth]{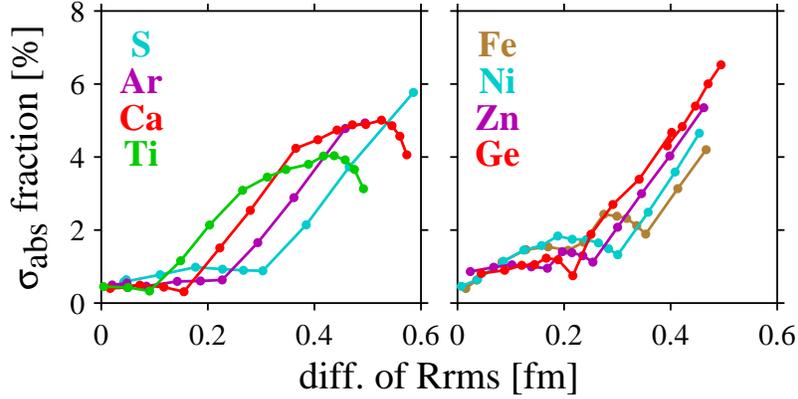}
  \caption{Calculated PDR strength fraction as a function of the neutron
skin thickness.
}
  \label{PDRcorr}
\end{figure}
Finally, let us examine the correlation between $f_{\rm PDR}$ and the neutron
skin thickness.
The skin thickness is defined by the difference in radius between
neutrons and protons.
Plotting the PDR fraction as a function of the skin thickness,
we observe a linear correlation between them, but only in specific regions
of the neutron number.
This is illustrated in Fig. \ref{PDRcorr} for isotopes with $Z=16--22$
and with $Z=26--32$, which show the kinks at $N=28$ and 50.
The PDR fraction in each isotopic chain shows a linear correlation with
the skin thickness in the regions of the neutron number
$N=28-34$ and $N \ge 50$.
The positions of the kinks are located at different values of
the skin thickness for different isotopes.
However, the slope is universal for all the isotopes;
$0.18\sim 0.20$ fm$^{-1}$.
Despite the fact that the deformation and shell ordering are different and
vary from nucleus to nucleus,
the universal linear correlation remains valid for $50\leq N < 76$.
It should be noted that the linear correlation can be observed
 only for each isotopic chain.
Deleting the lines connecting isotopic chains in Fig. \ref{PDRcorr},
we only see scattered points showing a weak correlation.
Again, for detailed analysis on this issue, readers are referred to
Ref.~\cite{INY11}.


\begin{theacknowledgments}
The work is supported by KAKENHI (Nos. 21340073 and 20105003) and
by SPIRE, MEXT, Japan.
The numerical calculations were performed on
PACS-CS and T2K supercomputers in University of Tsukuba,
on Hitachi SR11000 at KEK,
and on the RIKEN Integrated Cluster of Clusters (RICC).
\end{theacknowledgments}



\bibliographystyle{aipproc}   

\bibliography{myself,nuclear_physics,current,chemical_physics}

\IfFileExists{\jobname.bbl}{}
 {\typeout{}
  \typeout{******************************************}
  \typeout{** Please run "bibtex \jobname" to optain}
  \typeout{** the bibliography and then re-run LaTeX}
  \typeout{** twice to fix the references!}
  \typeout{******************************************}
  \typeout{}
 }

\end{document}